\documentclass[a4paper]{article}
\usepackage{CNVSRC2023}
\usepackage{epsfig,amssymb,amsmath,graphicx}
\usepackage{booktabs}
\usepackage{multirow}
\usepackage{amssymb}
\usepackage{pifont}
\usepackage{hyperref}

\ninept

\title{The NPU-ASLP-LiAuto System Description for Visual Speech Recognition in CNVSRC 2023}

\name{He Wang$^1$, Pengcheng Guo$^1$, Wei Chen$^2$, Pan Zhou$^2$, Lei Xie$^{1*}$}

\address{$^1$Audio, Speech and Language Processing Group (ASLP@NPU), School of Computer Science,\\Northwestern Polytechnical University, Xian, China\\$^2$Space AI, Li Auto  \\
{\small \tt hwang2001@mail.nwpu.edu.cn}}%

\providecommand{\keywords}[1]{\textbf{\textit{Index terms---}} #1}
\begin{document}
\maketitle

\begin{abstract}
This paper delineates the visual speech recognition (VSR) system introduced by the NPU-ASLP-LiAuto (Team 237) in the first \textbf{C}hi\textbf{n}ese Continuous \textbf{V}isual \textbf{S}peech \textbf{R}ecognition \textbf{C}hallenge (CNVSRC) 2023, engaging in the fixed and open tracks of Single-Speaker VSR Task, and the open track of Multi-Speaker VSR Task.
In terms of data processing, we leverage the lip motion extractor from the baseline\footnote{https://github.com/MKT-Dataoceanai/CNVSRC2023Baseline} to produce multi-scale video data.
Besides, various augmentation techniques are applied during training, encompassing speed perturbation, random rotation, horizontal flipping, and color transformation.
The VSR model adopts an end-to-end architecture with joint CTC/attention loss, comprising a ResNet3D visual frontend, an E-Branchformer encoder, and a Transformer decoder. 
Experiments show that our system achieves 34.76\% CER for the Single-Speaker Task and 41.06\% CER for the Multi-Speaker Task after multi-system fusion, ranking first place in all three tracks we participate.
\end{abstract}


\section{Introduction}
Leveraging the strides in deep learning, automatic speech recognition (ASR) has made substantial advancements, reaching parity with human performance on certain benchmarks. 
Nevertheless, not all scenarios boast access to high-quality speech audio. 
In response to the challenges posed by such scenarios, there has been a surge of interest among researchers in visual speech recognition (VSR).


To further advance research in VSR, the first \textbf{C}hi\textbf{n}ese Continuous \textbf{V}isual \textbf{S}peech \textbf{R}ecognition \textbf{C}hallenge\footnote{http://cnceleb.org/competition} (CNVSRC) 2023 is initiated, aiming to probe the performance of large vocabulary continuous visual speech recognition (LVCVSR) in two scenarios: reading in a recording studio and speech on the Internet. 
Furthermore, CNVSRC 2023 employs the CN-CVS~\cite{chen2023cn} dataset as its training set and defines two tasks: Single-Speaker VSR Task (T1) and Multi-Speaker VSR Task (T2). 
Independent development and evaluation sets accompany each task, denoted as CNVSRC-Single.Dev/Eval (T1.Dev/Eval) and CNVSRC-Multi.Dev/Eval (T2.Dev/Eval), respectively. 
Moreover, each task includes a fixed track, which restricts the use of data solely to that released by the challenge, and an open track without constraints on data usage.

This study describes our system in CNVSRC 2023.
As for the data processing, we extract lip motion video data by crop sizes of 48, 64, 96, and 112 to build multi-scale systems.
The VSR model adopts an end-to-end architecture with joint CTC/attention loss, consisting of dynamic augmentation, visual frontend, encoder, and decoder. 
Moreover, we diversify our systems by incorporating distinct encoders, including recently proposed E-Branchformer~\cite{kim2023branchformer}, Branchformer~\cite{peng2022branchformer}, and Conformer~\cite{gulati2020conformer}.
After using Recognizer Output Voting Error Reduction~\cite{fiscus1997post} (ROVER) for post-fusion, we attain CERs of 34.47\% and 34.76\% on the T1.Dev and T1.Eval datasets, alongside 41.39\% and 41.06\% on the T2.Dev and T2.Eval datasets. 
Our systems rank first place in the open tracks of both tasks and the fixed track of the Single-Speaker VSR Task.

\section{Proposed system}
\begin{figure}[t]
  \centering
  \centerline{\includegraphics[width=8.0cm]{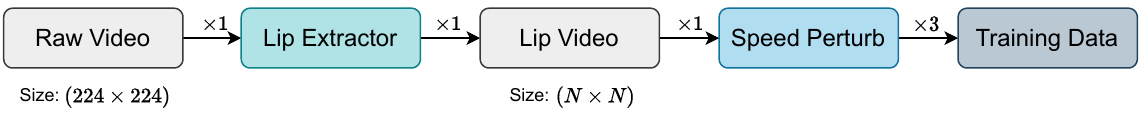}}
\caption{The flow chart of data processing. $\times$ n refers to n times the data of CN-CVS and CNVSRC-Single/Multi datasets. (N $\times$ N) refers to multi-scale video data.}
\label{fig:1}
\end{figure}
\subsection{Data processing}
Fig. \ref{fig:1} shows the data processing flow of our system. 
The initial training videos, sized 224 $\times$ 224, are processed via the lip extractor to obtain N $\times$ N (where N can be 48, 64, 96, or 112) sized lip motion video data.
Subsequently, the moivepy\footnote{https://pypi.org/project/moviepy} tool is employed to apply speed perturbation with rates of 0.9, 1.0, and 1.1, which results in three times training data in total.

\begin{figure}[t]
  \centering
  \centerline{\includegraphics[width=8.0cm]{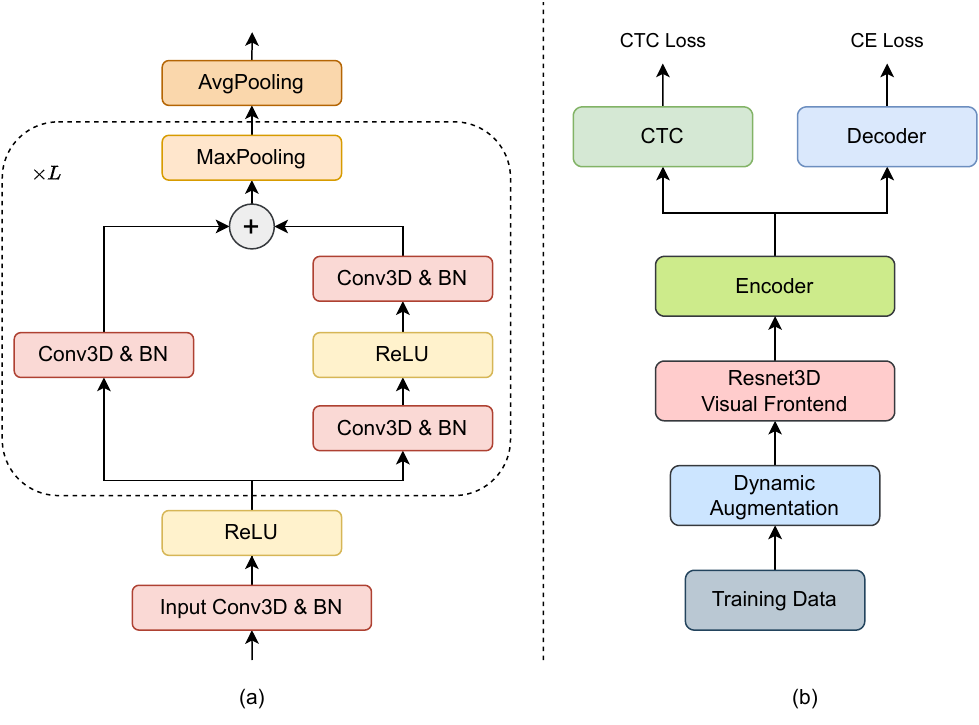}}
\caption{(a) Detail structure of the ResNet3D Visual Frontend; (b) System overview of our VSR system.}
\label{fig:2}
\end{figure}

\subsection{Visual speech recognition}
Fig. \ref{fig:2}(b) shows the overall architecture of our VSR system, comprising four main components: dynamic augmentation, visual frontend, encoder, and decoder. 
The training video data is first augmented by random rotation, horizontal flipping, and color transformation through the dynamic augmentation module implemented by the kornia\footnote{https://github.com/kornia/kornia} tool. 
Subsequently, it passes through the ResNet3D Visual Frontend module for visual feature extraction, with detailed information shown in Fig. \ref{fig:2}(a).

The design of the visual frontend module draws inspiration from the classic ResNet, where 2D convolutions are replaced with 3D ones. 
To reduce the computational cost, the network structure is simplified. 
Specifically, the dynamically augmented video data passes through a 3D convolution input layer to project the channel to a higher dimension. 
It then goes through $L$ (where $L=5$) ResNet blocks. 
In each block, the input visual features are modeled by independent convolution modules to achieve a higher and equal dimension of channels before the residual connection. 
Subsequently, MaxPooling is applied to perform two-fold downsampling on the height and width dimensions. 
Finally, AvgPooling is used to average the height and width dimensions, completing the process of visual dimension reduction and feature extraction.

For the encoder of the VSR system, we adopt the recently proposed E-Branchformer. 
Moreover, we build diverse VSR systems featuring different encoders such as Branchformer and Conformer, facilitating multi-system fusion. 
Regarding the decoder, a common Transformer decoder is utilized. 
The final model loss is composed of the CTC loss derived from the encoder and the cross-entropy (CE) loss computed on the decoder.

\section{Experiment}

\subsection{Setup}
All systems are implemented with the ESPnet\footnote{https://github.com/espnet/espnet} toolkit, only using the CN-CVS and the development set of CNVSRC-Single/Multi datasets.
For the encoder of the VSR system, we use a 12-layer E-Branchformer block, each with 256 attention units, 4 attention heads, and 1024 feed-forward units. 
Additionally, the decoder contains 6 Transformer layers, each with 4 attention heads and 2048 feed-forward units. 
The visual frontend is a 5-layer ResNet3D module, whose channels are 32, 64, 64, 128, 256, and kernel size is 3. 
To achieve better results by post-fusion, we also build Conformer and Branchformer VSR systems with similar param sizes to the E-Branchformer one.

\subsection{Inference Procedure}
We build 24-layer Transformer language models (LM) using the text from CN-CVS and the CNVSRC-Single/Multi datasets for the shallow fusion in Single-Speaker and Multi-Speaker Tasks, respectively. 
The embedding and attention dimensions are set to 512, with 8 attention heads and 2048 feed-forward units. 
During decoding, the beam size is set to 48 and both CTC and LM contribute to scoring with weights of 0.5 and 0.4, respectively.

\begin{table}[t]
	\centering
	\caption{The CER(\%) results of our VSR systems on Dev and Eval sets in T1 and T2 tasks. 
    Conf, Branch and E-Branch represent Conformer, Branchformer, and E-Branchformer encoder, respectively. 
    Crop refers to the size of training lip motion video data. SP means speed perturbation is applied or not.}
    \resizebox{\linewidth}{!}{
        \begin{tabular}{cccccccc}
    		\toprule
    		System & Encoder & Crop & SP & T1.Dev & T1.Eval & T2.Dev & T2.Eval \\
    		\hline
            Baseline\footnote{https://github.com/MKT-Dataoceanai/CNVSRC2023Baseline} & Conf & 96 & \text{\ding{55}} & 48.57 & 48.60 & 58.77 & 58.37 \\
    		M1 & Conf & 96 & \text{\ding{51}} & 39.43 & 39.99 & 46.08 & 45.73 \\
    		M2 & Branch & 96 & \text{\ding{51}} & 39.00 & 39.36 & 46.63 & 46.37 \\
    		M3 & E-Branch & 96 & \text{\ding{51}} & 38.59 & 38.61 & 46.26 & 45.80 \\
            M4 & E-Branch & 48 & \text{\ding{55}} & 46.88 & 45.81 & 55.58 & 55.51 \\
            M5 & E-Branch & 64 & \text{\ding{55}} & 44.40 & 43.59 & 53.64 & 52.98 \\
            M6 & E-Branch & 80 & \text{\ding{55}} & 42.95 & 42.26 & 50.77 & 50.38 \\
            M7 & E-Branch & 96 & \text{\ding{55}} & 40.56 & 40.42 & 47.16 & 46.53 \\
            M8 & E-Branch & 112 & \text{\ding{55}} & 38.46 & 38.95 & 45.17 & 44.87 \\            
            ROVER & - & - & - & \textbf{34.47} & \textbf{34.76} & \textbf{41.39} & \textbf{41.06} \\
    		\bottomrule
    	\end{tabular}
    }
	\label{table-1}
\end{table}
\subsection{Results}
Table \ref{table-1} shows the outcomes of our VSR systems (M1$\sim$M8) and the system fusion result obtained by the ROVER. 
Specifically, in terms of the encoder, the performance of the E-Branchformer surpasses that of the Branchformer and Conformer (M1$\sim$M3). 
Moreover, as the crop size increases from 48 to 112, the performance of our VSR system improves (M4$\sim$M8).
Using three times speed perturbation also yields a certain improvement in CER (M3 and M7). 
After multi-system fusion by ROVER, we achieve 34.47\% and 34.76\% CERs on the Dev and Eval sets of T1, alongside 41.39\% and 41.06\% CERs of T2, respectively.

\section{Conclusion}
This paper describes the visual speech recognition (VSR) system introduced by the NPU-ASLP-LiAuto (Team 237) in the \textbf{C}hi\textbf{n}ese Continuous \textbf{V}isual \textbf{S}peech \textbf{R}ecognition \textbf{C}hallenge (CNVSRC) 2023, engaging in the fixed and open tracks of Single-Speaker VSR Task, and the open track of Multi-Speaker VSR Task.
In detail, we extract lip motion video at multiple scales and employ different mainstream encoders to build diverse VSR systems for multi-system fusion. 
Finally, we achieve CERs of 34.76\% and 41.06\% on the final evaluation sets of Single-Speaker VSR Task and Multi-Speaker VSR Task, respectively, ranking first place in all three tracks we participate.

%

%

\bibliographystyle{IEEEbib}
\bibliography{BibEntries}

\end{document}